\documentclass[preprint]{aastex}

\begin{document}

\title {$\gamma$ Doradus Pulsations in the Eclipsing Binary Star KIC 6048106 }
\author{Jae Woo Lee$^{1,2}$}
\affil{$^1$Korea Astronomy and Space Science Institute, Daejeon 34113, Korea}
\affil{$^2$Astronomy and Space Science Major, Korea University of Science and Technology, Daejeon 34113, Korea}
\email{jwlee@kasi.re.kr}

\begin{abstract}
We present the ${\it Kepler}$ photometry of KIC 6048106 exhibiting O'Connell effect and multiperiodic pulsations. Including 
a starspot on either of the components, light-curve synthesis indicates that this system is a semi-detached Algol with 
a mass ratio of 0.211, an orbital inclination of 73.9 deg, and a large temperature difference of 2,534 K. To examine in detail 
both spot variations and pulsations, we separately analyzed the {\it Kepler} time-series data at the interval of an orbital period 
by an iterative way. The results reveal that the variable asymmetries of the light maxima can be interpreted as the changes of 
a magnetic cool spot on the secondary component with time. Multiple frequency analyses were performed in the outside-eclipse light 
residuals after removal of the binarity effects from the observed {\it Kepler} data. We detected 30 frequencies with signal to 
noise amplitude ratios larger than 4.0, of which six ($f_2$--$f_6$ and $f_{10}$) can be identified as high-order (17 $\le n \le$ 25) 
low-degree ($\ell$ = 2) gravity-mode pulsations that were stable during the observing run of 200 d. In contrast, the other frequencies 
may be harmonic and combination terms. For the six frequencies, the pulsation periods and pulsation constants are in the ranges of 
0.352$-$0.506 d and 0.232$-$0.333 d, respectively. These values and the position on the HR diagram demonstrate that the primary star 
is a $\gamma$ Dor variable. The evolutionary status and the pulsation nature of KIC 6048106 are discussed. 
\end{abstract}

{}

\section{INTRODUCTION}

$\gamma$ Dor stars are A$-$F stars of luminosity class IV$-$V near the red edge of the $\delta$ Sct instability strip in 
the Hertzsprung-Russell (HR) diagram. Their observational properties are very similar to those of $\delta$ Sct stars, but they pulsate 
in high-order gravity ($g$) modes driven by convective blocking (Guzik et al. 2000; Dupret et al. 2004, 2005) with typical periods 
of 0.4$-$3 d and pulsation constants of $Q >$ 0.23 d (Kaye et al. 1999; Henry et al. 2005). These pulsating stars are of great interest 
to asteroseismic studies because the $g$ modes assist in probing the deep stellar interiors near the core region. The number of 
the $\gamma$ Dor-type stars has increased dramatically by space-based missions such as {\it CoRot} (Hareter et al. 2010) and 
{\it Kepler} (Balona et al. 2011; Bradley et al. 2015). Nonetheless, only thirteen eclipsing binaries (EBs) have been known 
to contain $\gamma$ Dor pulsating candidates (Maceroni et al. 2014; Kurtz et al. 2015; \c Cakirli \& Ibano\v{g}lu 2016), of which 
V551 Aur may be a $\delta$ Sct star as indicated by its pulsation frequencies (Liu et al. 2012).

Because EBs are a primary source of the fundamental stellar properties such as mass and radius, the pulsating EBs are ideal 
targets for the study of the interior structure and evolution of stars from their binarity and pulsation features. Recently, 
\c Cakirli \& Ibano\v{g}lu (2016) suggested three possible relationships between the pulsation periods for $\gamma$ Dor components 
and other parameters (binary orbital periods for 11 EBs, surface gravities for 6 pulsating components, and the gravitational forces 
from companions for 5 EBs). As in case of the EBs with $\delta$ Sct components (Soydugan et al. 2006; Liakos et al. 2012), 
the longer the binary orbital period, the longer the pulsation period: $P_{\rm pul}$ = 0.425$P_{\rm orb}$ $-$ 0.355. In addition, 
as the surface gravity of pulsating components and the gravitational force from the companions decrease, their pulsation period 
increases. However, these relationships need further confirmation by new discoveries because the number of such stars remains small. 

KIC 6048106 (R.A.$_{2000}$=19$^{\rm h}$34$^{\rm m}$14$\fs028$; decl.$_{2000}$=+41$^{\circ}$23${\rm '}$43$\farcs$26; $K_{\rm p}$=$+$14.091; $g$=$+$14.303; $g-r$=$+$0.283) 
was announced to be an EB pulsating at frequencies of 0.43$-$3.28 d$^{-1}$ by Gaulme \& Guzik (2014). In this paper, we demonstrate 
that the binary system is a semi-detached Algol with a $\gamma$ Dor-type pulsating component, based on the precise and nearly-continuous 
{\it Kepler} data during a period of approximately 200 d. In Section 2, we carry out light-curve synthesis and present the binary parameters 
including the absolute dimensions. Section 3 describes the frequency analysis for the light residuals from each spot model. Finally, 
we summarize and discuss our conclusion in Section 4.

\section{LIGHT-CURVE SYNTHESIS AND ABSOLUTE DIMENSIONS}

KIC 6048106 was observed in a long cadence (LC) mode of 29.42 min during Quarters 14 and 15 by the {\it Kepler} satellite. 
The contamination level of the observations by nearby stars is estimated to be 0.028, where a value of 0 implies no contamination 
and 1 implies all background. For this study, we used the simple aperture photometry (SAP) data detrended in the {\it Kepler} EB 
catalogue (Pr\v sa et al. 2011; Slawson et al. 2011)\footnote{http://keplerebs.villanova.edu/}. The {\it Kepler} data of 
KIC 6048106 are depicted in the top panel of Figure 1 as the normalized flux versus the orbital phase. The light curve is superficially 
similar to that of Algol and displays the inverse O'Connell effect whereby Max I (following primary eclipse) is fainter than Max II 
by about 0.005 mag. In short-period Algols, this effect is usually interpreted to be due to magnetic dynamo-related activity 
in a cool companion and/or mass transfer between both components if a binary system is in a semi-detached configuration 
(e.g., KIC 4739791, Lee et al. 2016b). 

In order to understand the physical properties of this system, we analyzed simultaneously all {\it Kepler} data in a manner almost 
identical to that for the pulsating EBs V404 Lyr (Lee et al. 2014) and KIC 6220497 (Lee et al. 2016a) by using the 2007 version of 
the Wilson-Devinney binary code (Wilson \& Devinney 1971; van Hamme \& Wilson 2007; hereafter W-D) and the so-called $q$-search 
procedure. In our synthesis, the mean light level at phase 0.75 was set to unity and the surface temperature of the hotter and 
more massive star was initialized at $T_{1}$=7,000$\pm$245 K from the revised {\it Kepler} Input Catalogue (KIC) properties 
(Huber et al. 2014). Adjustable parameters include the orbital ephemeris ($T_0$ and $P$), the mass ratio ($q$), 
the orbital inclination ($i$), the effective temperatures ($T_{1,2}$) and the dimensionless surface potentials ($\Omega_{1,2}$) of 
the components, and the monochromatic luminosity ($L_1$). Throughout our analyses, a synchronous rotation for both components and 
a circular orbit were adopted and the detailed reflection effect was used. In this paper, we refer to the primary and secondary stars 
as those being eclipsed at Min I (phase 0.0) and Min II, respectively.

First, because there is no light-curve solution for KIC 6048106, we conducted an extensive $q$-search for various modes 
of the W-D code to obtain binary parameters. The behavior of the weighted sum of the squared residuals ($\sum W(O-C)^2$; 
hereafter $\sum$) was used to estimate the potential reality of each model. This procedure showed acceptable photometric solutions 
only for the semi-detached mode for which the less massive secondary star accurately fills its inner Roche lobe. As displayed in 
Figure 2, the $q$-search results indicate that the optimal solution is around $q$=0.21, which was adopted as the initial value and 
adjusted to derive all subsequent photometric solutions. The unspotted solution is listed in the second and third columns of Table 1 
and appears as a dashed curve in the top panel of Figure 1. The light residuals from this solution are plotted in the middle panel 
of Figure 1, wherein it can be seen that the unspotted model does not describe the observed {\it Kepler} data satisfactorily. 
In this and subsequent syntheses, the errors for the adjustable parameters were established following the procedure described by 
Koo et al. (2014). For this procedure, the observed {\it Kepler} data were split into ten subsets. 

The light asymmetries in KIC 6048106 may be attributed to either a magnetic cool spot located on the surface of the late-type 
secondary star or a hot spot on the detached primary as a result of the impact of the gas stream from the lobe-filling companion 
by mass transfer. The two possible spot models were applied to fit the asymmetrical light curve, and the results are given in 
columns (4)--(7) of Table 1. The synthetic curve from the cool-spot model is displayed as the solid curve in the top panel of 
Figure 1, and the residuals from the spot model are plotted in the bottom panel. From the table and figure, we can see that 
the spot models improve the light-curve fitting and give smaller values of $\sum$ than does the unspotted model. Although there 
are no $\sum$ differences between the two spot models, it may be more reasonable to regard the main cause of the spot activity 
to be a magnetized cool spot on the secondary star, because the {\it Kepler} data showed the variable asymmetries of 
the light maxima with time.

Considering its error, the effective temperature of the primary component corresponds to a normal main-sequence star with 
a spectral type of approximately F0$\pm$1. We assumed the primary's mass to be $M_1$=1.50$\pm$0.08$M_\odot$ from 
Harmanec's (1988) established relationship between spectral type and stellar mass. The absolute dimensions for KIC 6048106 can be 
roughly estimated from our photometric solutions and $M_1$. These are given in the bottom of Table 1, where the luminosity ($L$) 
and bolometric magnitudes ($M_{\rm bol}$) were computed by adopting $T_{\rm eff}$$_\odot$=5,780 K and $M_{\rm bol}$$_\odot$=+4.73 
for solar values. For the absolute visual magnitudes ($M_{\rm V}$), we used the bolometric corrections (BCs) appropriate for 
the temperature of each component from the expression between $\log T_{\rm eff}$ and BC given by Torres (2010).

\section{STARSPOT ACTIVITY AND PULSATIONAL CHARACTERISTICS}

Time-series data of 20 d for the {\it Kepler} observations are displayed as cyan circles in Figure 3. As shown in the figure, 
the brightness of KIC 6048106 has varied visibly with time, which might be a result of spot variations and pulsations. In order 
to study in detail the spot behavior and obtain more reliable frequency analyses, we divided the light curve of KIC 6048106 
into 113 segments at the interval of an orbital period and separately analyzed them. Firstly, we obtained the light residuals 
after subtracting the cool-spot model in Table 1 from the observed {\it Kepler} data and applied a multiple frequency analysis 
to only the outside-eclipse residuals using the PERIOD04 program (Lenz \& Breger 2005). The frequency analysis was performed 
on the range from 0 to the Nyquist limit of $f_{\rm Ny}$=24.47 d$^{-1}$. As in the case of V404 Lyr (Lee et al. 2014), 
the successive prewhitening of each frequency peak was carried out by applying the light residuals to the equation of 
$Z$ = $Z_0$ + $\Sigma _{i}$ $A_i \sin$(2$\pi f_i t + \phi _i$). Here, $Z$ and $Z_0$ denote the calculated magnitude and zero point, 
respectively, $A_i$ and $\phi _i$ are the amplitude and phase of the $i$th frequency, respectively, and $t$ is the time of 
each measurement. We detected the frequencies with signal to noise amplitude (S/N) ratios larger than 4.0 (Breger et al. 1993). 
Secondly, we removed the pulsation signatures from the observed data, and each pulsation-subtracted light curve was solved by using 
the cool-spot model parameters as initial values and adjusting the epoch and spot parameters, with the exception of the colatitude. 
Thirdly, we removed the spot solution from each light curve in the original data and performed frequency analysis for the entire set
of residuals, excluding the data around the primary and secondary minima. 

This procedure was repeated 5 times until the detected frequencies were unchanged. In the first iteration, very low frequency 
signals such as 0.00635 d$^{-1}$ and 0.03706 d$^{-1}$ were detected, but they disappeared in subsequent analyses. The final epochs 
and spot parameters are listed in Table 2, and the pulsation-subtracted data and spot models for each light curve are plotted 
as plus symbols and solid curves, respectively, in Figure 3. The light residuals from the detailed analysis are displayed in 
the upper panel of Figure 4 as magnitude versus BJD, wherein the lower panel presents a short section of the residuals. 
In Figure 5, we showed the amplitude spectra for KIC 6048106 before and after prewhitening the first 6 frequencies and all 30 
frequencies. The synthetic curve obtained from the 30-frequency fit is displayed in the lower panel of Figure 4. The result for 
the multiple frequency analysis is given in Table 3, where the frequencies are listed in order of detection, and the noises are 
calculated in a range of 5 d$^{-1}$ around each frequency. We can see that all signals lie in the low-frequency $g$-mode region. 
The uncertainties in the table were derived according to Kallinger et al. (2008). 

In order to examine the frequency variations with time, we analyze the out-of-eclipse residuals at intervals of $\sim$50 d. 
The results from the analyses are not much different from each other, and more than 11 frequencies were detected at each subset 
with the same criterion of S/N$>$4.0. We plotted the stability of the 8 frequencies detected repeatedly in Figure 6. Among these, 
the frequencies of $f_2, f_3, f_4, f_5, f_6$, and $f_{10}$ were almost constant with time, while the $f_1$ and $f_{16}$ 
frequencies varied significantly. The unstable frequencies might result from alias effects caused by the orbital frequency 
($f_{\rm orb}$ = 0.64129 d$^{-1}$). Within the frequency resolution of 0.008 d$^{-1}$ (Loumos \& Deeming 1978), we searched 
for possible harmonic and combination frequencies. As listed in the last column of Table 3, six ($f_2-$$f_6$, $f_{10}$) are 
pulsation frequencies, and $f_9$ and $f_{15}$ appear to be the sidelobes split from the orbital frequency $f_{\rm orb}$ by 
$\sim$0.011 d$^{-1}$. The other frequencies may come from orbital harmonics and combination frequencies, some of which could be 
partially attributed to imperfect removal of the binary effects in the observed light curve.

The {\it Kepler} LC data integrate over 270 exposures with a sampling rate of 6.54 s (including readout) to form 29.42 min 
observations. The merging effect reduces the amplitude of peaks in the power spectrum, which increases with higher frequencies 
(Murphy 2012; Lee et al. 2016b). On the other hand, some signals near the Nyquist frequency can be the reflections of real frequencies 
(2$f_{\rm Ny}-f_i$) (Murphy et al. 2013). However, because KIC 6048106 exhibit much lower frequencies than the Nyquist limit, 
the possibility of the reflections in the detected frequencies seems very unlikely.

\section{DISCUSSION AND CONCLUSIONS}

In this paper, we studied both the binarity and pulsation of KIC 6048106 from detailed analyses of the {\it Kepler} observations 
obtained during Quarters 14 and 15. The {\it Kepler} time-series data display mutiperiodic pulsations and the O'Connell effect 
with unequal light levels at the quadratures (Max I and Max II). The asymmetric light curve was modelled by applying a single spot 
to either of the components: a hot spot on the primary star and a cool spot on the secondary. Our light-curve synthesis indicates 
that KIC 6048106 is a classical Algol-type system with parameters of $q$=0.211, $i$=73$^\circ$.9, and ($T_{1}$--$T_{2}$)=2,534 K; 
the primary component fills about 52\% of its limiting lobe and is slightly smaller than the lobe-filling secondary. 
The locations of the components in the HR diagram are shown in Figure 7, together with those of other well-studied semi-detached 
Algols (\. Ibano\v{g}lu et al. 2006) and the $\gamma$ Dor stars in 9 EBs (Maceroni et al. 2014; \c Cakirli \& Ibano\v{g}lu 2016). 
Here, the dashed and dash-dotted lines are the instability strips of $\gamma$ Dor and $\delta$ Sct stars, respectively. 
The pulsating primary star of KIC 6048106 resides within the $\gamma$ Dor region on the zero-age main sequence (ZAMS), 
and the secondary lies in a location where the secondary components of other Algols exist. 

The {\it Kepler} data indicated that the light curve of KIC 6048106 has varied due to the combination of both spot and pulsation. 
To explore in detail the light variations, we individually analyzed the {\it Kepler} light curve at the interval of an orbital period 
by the iterative method described in the previous section. The variable asymmetries of the light maxima can be explained by 
the changes of the cool spot with time, which may be formed from magnetic dynamo-related activity because the system is rotating 
rapidly and the secondary component should be a deep convective envelope as surmised from its temperature. In order to understand 
the pulsational characteristics of the system, multiple frequency analyses were applied to the whole outside-eclipse light residuals, 
removing the binarity effects from the observed {\it Kepler} data. Thirty frequencies with S/N ratios larger than 4.0 were found 
in the range of 0.31$-$5.78 d$^{-1}$ with amplitudes between 0.14 and 3.29 mmag. Among these, six ($f_2, f_3, f_4, f_5, f_6$, and 
$f_{10}$) may be pulsation frequencies in $g$-mode region, which were stable during the observational interval of about 200 d. 
We computed the pulsation constants from the cool-spot model parameters in Table 1 and the well-known relation of 
$\log Q_i = -\log f_i + 0.5 \log g + 0.1M_{\rm bol} + \log T_{\rm eff} - 6.456$ (Petersen \& J\o rgensen 1972). The results are 
listed in the third column of Table 4. The $Q$ values and the position of KIC 6048106 on the HR diagram demonstrate that 
the detached primary component would be a $\gamma$ Dor-type pulsating star. On the other hand, all but the six frequencies appear 
to be harmonic and combination terms, some of which might arise from starspot activity or from alias effects caused by the orbital 
frequency. The orbital harmonics ($f_7, f_8, f_{11}, f_{12}, f_{19}, f_{20}$) can be stellar pulsations excited by the tidal forces 
of the secondary component (Welsh et al. 2011; Hambleton et al. 2013; Lee et al. 2016a).

As binary stars are generally supposed to reach the synchronization before their semi-detached phases, the pulsating primary component 
of KIC 6048106 may have a synchronized rotation of approximately 50 km s$^{-1}$. Thus, we can make a possible identification of 
the radial order ($n$) and spherical degree ($\ell$) for the observed frequencies with the Frequency Ratio Method (Moya et al. 2005; Su\'arez et al. 2005), 
which is useful for $\gamma$ Dor stars with rotational velocities of $v \sin i \la$ 70 km s$^{-1}$ and at least three $g$-mode 
frequencies. Furthermore, it is possible to obtain the corresponding value for the integral of the Brunt-V\"ais\"al\"a frequency 
($\cal J$). Following the procedure described by Lee et al. (2014), we determined the model frequency ratios ($f_i$/$f_5$)$_{\rm model}$ 
best-fitted to the observed ratios ($f_i$/$f_5$)$_{\rm obs}$, and identified the pulsation modes of the six frequencies. 
As listed in Table 4, the $f_2$, $f_3$, $f_4$, $f_5$, $f_6$ and $f_{10}$ frequencies are identified as degree $\ell$ = 2 for 
radial orders of $n$ = 25, 22, 24, 17, 21, and 18, respectively. The observed average value of $\cal J_{\rm obs}$ = 742.2$\pm$5.3 
for the six frequencies is close to the theoretical integral of $\cal J_{\rm theo}$ $\approx$ 700 $\mu$Hz for a model of 
$\log$ $T_{\rm eff}$=3.845, 1.5 $M_\odot$, and [Fe/H]=0.0 in the $\cal J -$ $\log$ $T_{\rm eff}$ diagram given by Moya et al. (2005).

Classical Algols are semi-detached interacting systems in which one type of interaction is mass transfer from the lobe-filling 
secondary to the detached primary component via the inner Lagrange $L_1$ point. The semi-detached configuration of KIC 6048106 
permits some mass transfer between the component stars by means of a gas stream. Just as with the mass-accreting $\delta$ Sct 
components of semi-detached Algols (Mkrtichian et al. 2004; the so-called oEA stars), the secondary to primary mass transfer 
could at least be partly responsible for the $\gamma$ Dor-type oscillations detected in this paper. In addition, 
the pulsations may be influenced by the tidal and gravitation forces from the secondary component. As mentioned in the Introduction, 
\c Cakirli \& Ibano\v{g}lu (2016) presented three empirical relations for the $\gamma$ Dor stars in EBs, where the equation and 
figure between the pulsation period $\log P_{\rm pul}$ and the gravitational force $\log (F/M_1)$ are not consistent with 
each other. We think that their equation (9) is $\log (F/M_1) = -2.021 \log P_{\rm pul} +$ 2.093. The physical properties 
of KIC 6048106 match well these relationships of the orbital periods, the surface gravities, and the gravitational forces against 
the pulsation periods. However, because only thirteen stars, including KIC 6048106, have been identified as EBs containing 
$\gamma$ Dor-type components, additional discoveries and follow-up observations will help to reveal more accurate properties of 
the pulsating EBs.

\acknowledgments{ }
This paper includes data collected by the {\it Kepler} mission. {\it Kepler} was selected as the 10th mission of the Discovery Program. 
Funding for the {\it Kepler} mission is provided by the NASA Science Mission directorate. We have used the Simbad database maintained 
at CDS, Strasbourg, France. This work was supported by the KASI (Korea Astronomy and Space Science Institute) grant 2016-1-832-01.

\clearpage
\begin{figure}
\includegraphics[scale=0.85]{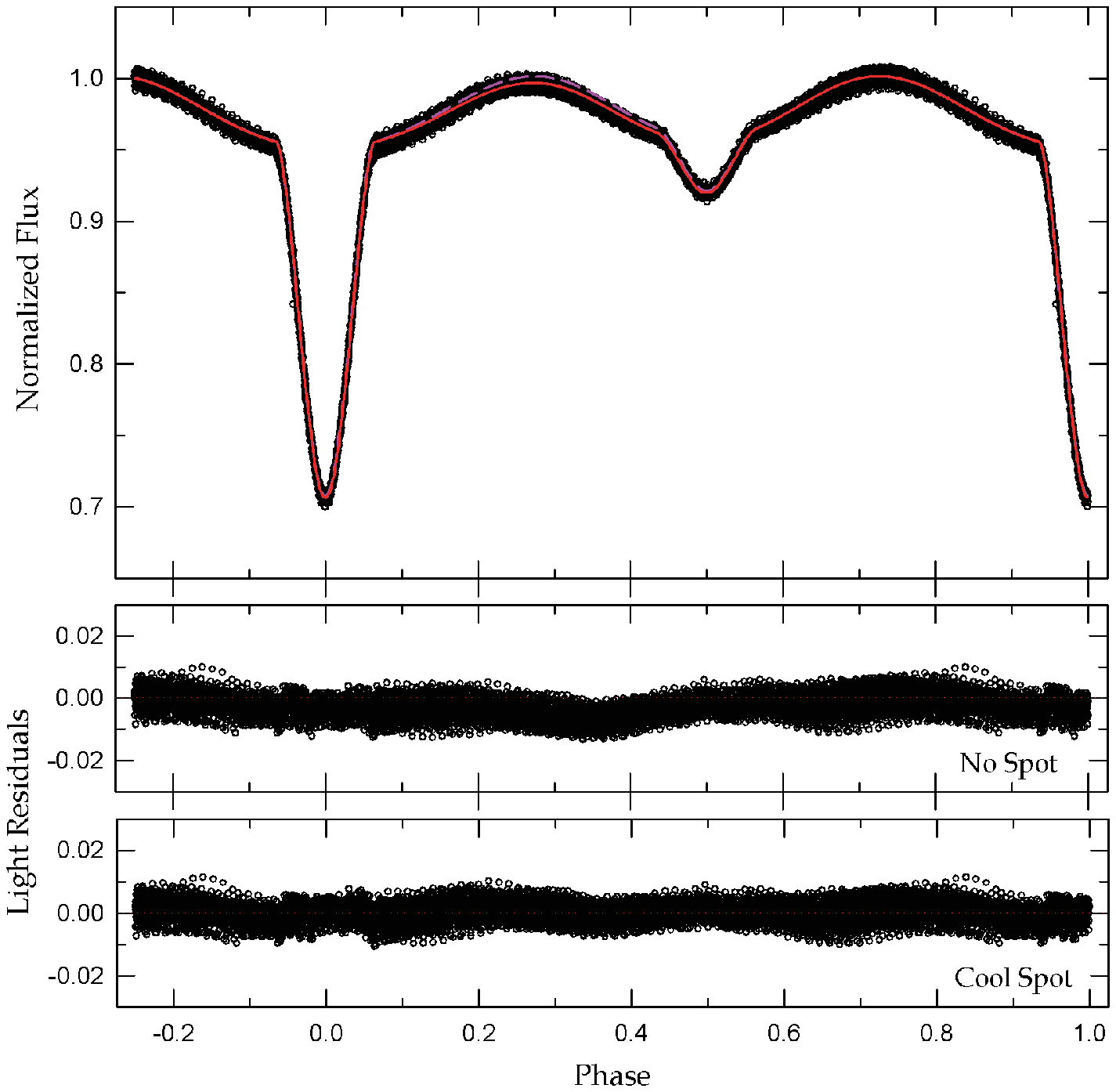}
\caption{Light curve of KIC 6048106 with the fitted models. In the top panel, the circles are individual measures from 
the {\it Kepler} spacecraft, and the dashed and solid curves represent the synthetic curves computed from no spot and 
the cool-spot model on the secondary star, respectively. The light residuals corresponding to the two models are plotted in 
the middle and bottom panels, respectively. }
\label{Fig1}
\end{figure}

\begin{figure}
\includegraphics[]{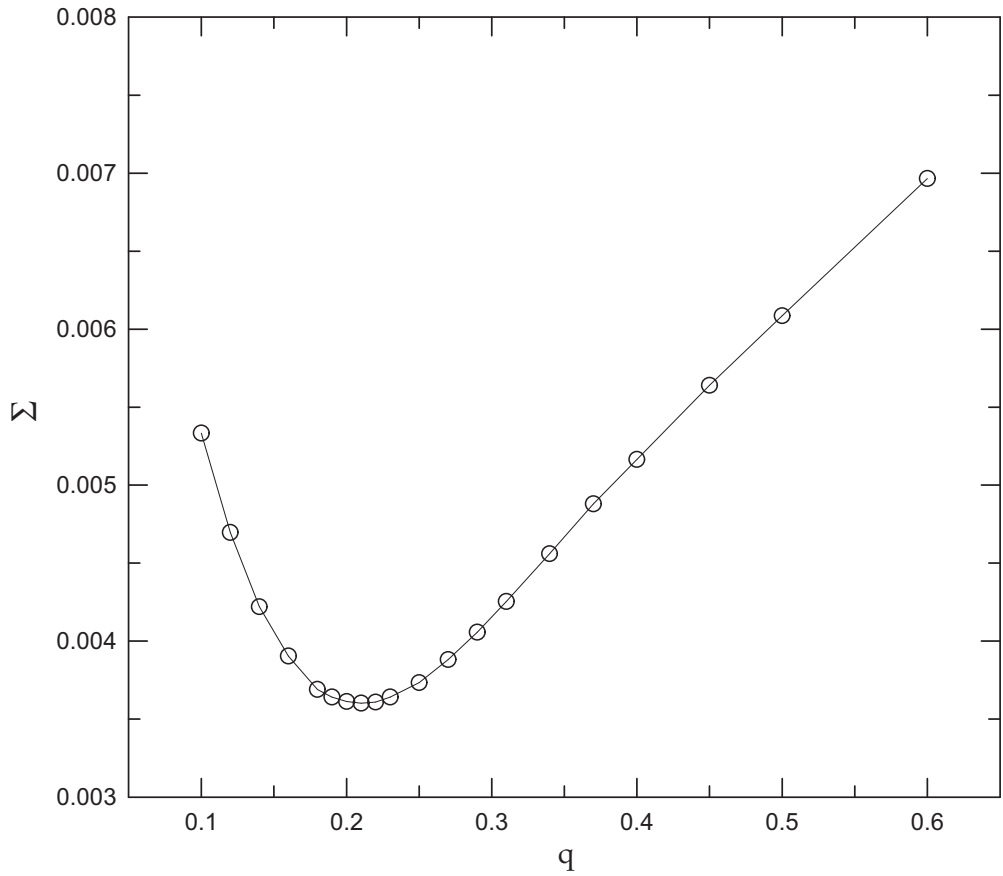}
\caption{Behavior of $\sum$ of KIC 6048106 as a function of mass ratio $q$, showing a minimum value at $q$=0.21. }
\label{Fig2}
\end{figure}

\begin{figure}
\includegraphics[]{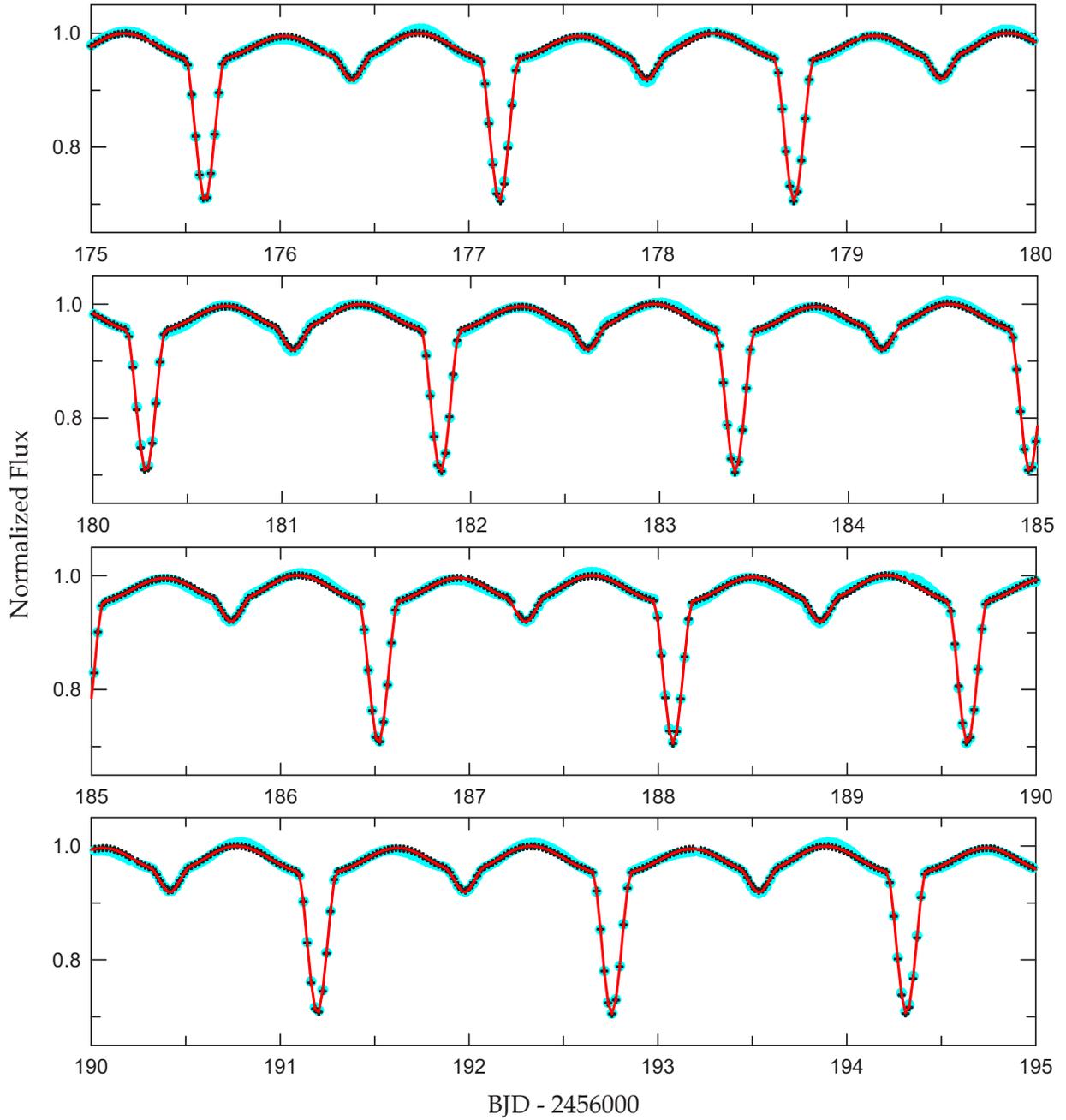}
\caption{Time-series data of 20 d for the observed (cyan circle) and pulsation-subtracted (black plus) light curves of KIC 6048106. 
The solid curves were obtained from each spot solution for the prewhitened data after removing pulsation signatures from 
the observed {\it Kepler} data. }
\label{Fig3}
\end{figure}

\begin{figure}
\includegraphics[]{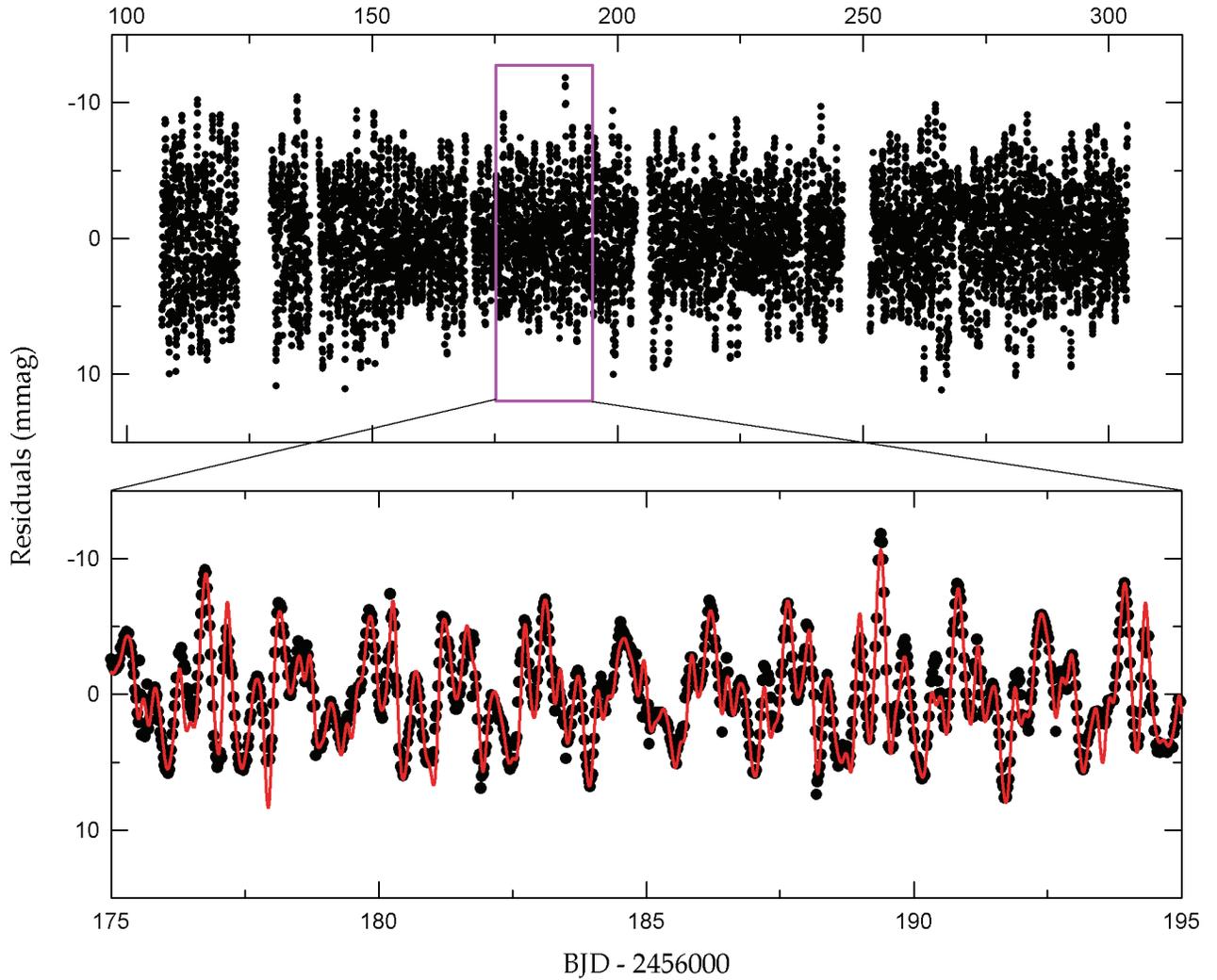}
\caption{Light residuals after subtracting the binarity effects from each light curve in the observed {\it Kepler} data. 
The lower panel presents a short section of the residuals marked using the inset box of the upper panel. The synthetic curve was 
computed from the 30-frequency fit to the entire data except for the times of both eclipses. }
\label{Fig4}
\end{figure}

\begin{figure}
\includegraphics[]{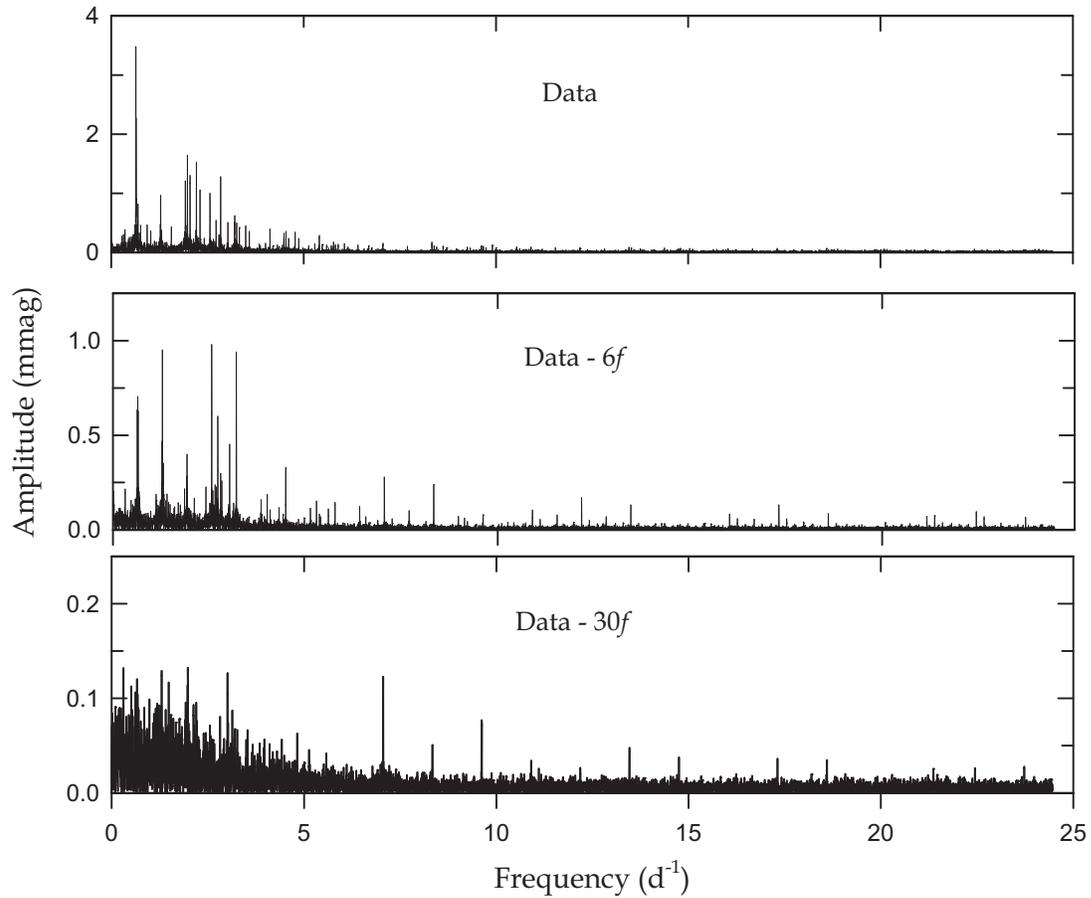}
\caption{Amplitude spectra before (top panel) and after pre-whitening the first 6 frequencies (middle panel) and all 30 frequencies 
(bottom panel) from the PERIOD04 program for the whole outside-eclipse residual data (orbital phases 0.065$-$0.435 and 0.565$-$0.935). }
\label{Fig5}
\end{figure}

\begin{figure}
\includegraphics[]{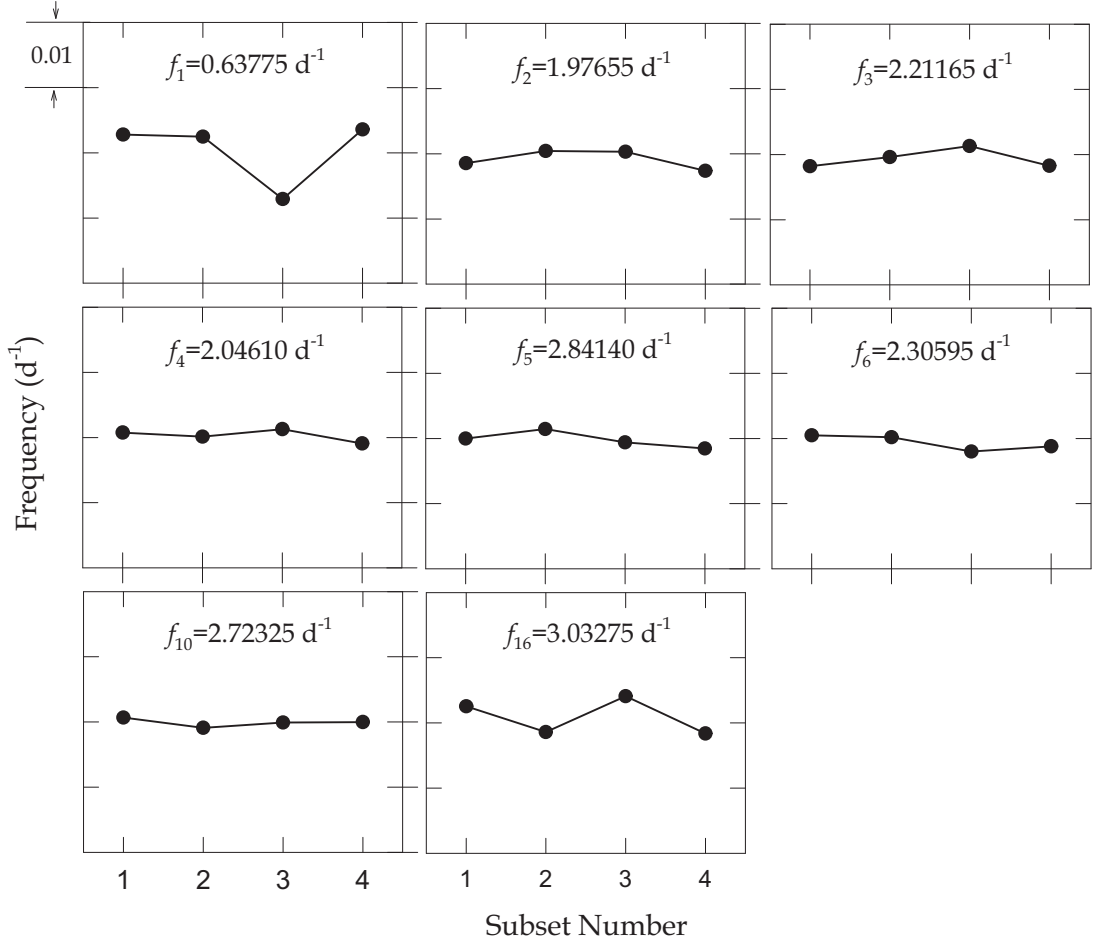}
\caption{Variability of the main frequencies detected in the four subsets at intervals of $\sim$50 d. In all panels, 
the y-axes are scaled to 0.04 d$^{-1}$, and the tick intervals are 0.01 d$^{-1}$. }
\label{Fig6}
\end{figure}

\begin{figure}
\includegraphics[]{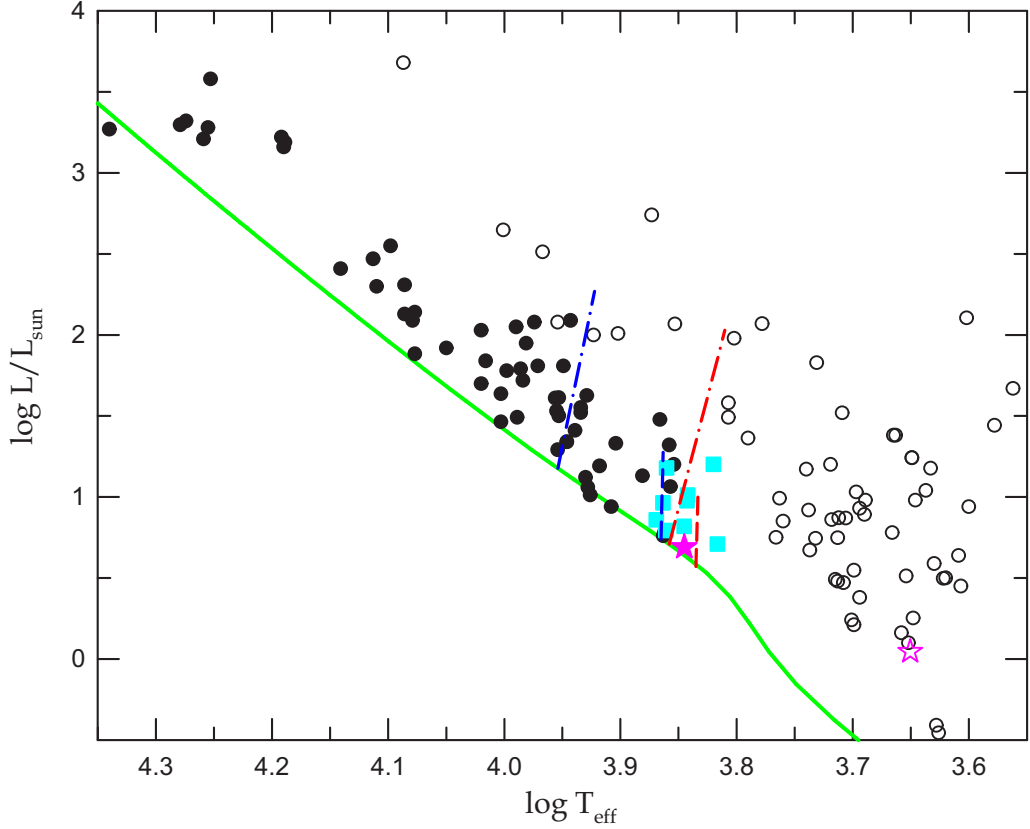}
\caption{Position on the HR diagrams for KIC 6048106 (star symbols) and other Algol EBs. The filled and open circles refer to 
the primary and secondary components of the semi-detached Algols, respectively, and the squares are 9 $\gamma$ Dor stars 
in EBs with known parameters. The solid line denotes the ZAMS for solar metallicity from Tout et al. (1996). The dashed and 
dash-dotted lines represent the instability strips of $\gamma$ Dor (Warner et al. 2003; \c Cakirli 2015) and $\delta$ Sct stars 
(Rolland et al. 2002; Soydugan et al. 2006), respectively. } 
\label{Fig7}
\end{figure}

\clearpage
\begin{deluxetable}{lcccccccc}
\tabletypesize{\small}  
\tablewidth{0pt} 
\tablecaption{Binary Parameters of KIC 6048106 Obtained by Fitting all Observed {\it Kepler} Data Simultaneously }
\tablehead{
\colhead{Parameter}                      & \multicolumn{2}{c}{Unspotted Model}         && \multicolumn{2}{c}{Hot-spot Model}          && \multicolumn{2}{c}{Cool-spot Model}         \\ [1.0mm] \cline{2-3} \cline{5-6} \cline{8-9} \\[-2.0ex]
                                         & \colhead{Primary} & \colhead{Secondary}     && \colhead{Primary} & \colhead{Secondary}     && \colhead{Primary} & \colhead{Secondary}                                                  
}
\startdata                                                                                                                                                                           
$T_0$ (BJD)                              & \multicolumn{2}{c}{2,456,108.55098(27)}     && \multicolumn{2}{c}{2,456,108.55062(25)}     && \multicolumn{2}{c}{2,456,108.55110(23)}     \\
$P$ (days)                               & \multicolumn{2}{c}{1.5593587(27)}           && \multicolumn{2}{c}{1.5593591(27)}           && \multicolumn{2}{c}{1.5593595(18)}           \\
$q$                                      & \multicolumn{2}{c}{0.2097(33)}              && \multicolumn{2}{c}{0.2096(35)}              && \multicolumn{2}{c}{0.2125(40)}              \\
$i$ (deg)                                & \multicolumn{2}{c}{73.899(67)}              && \multicolumn{2}{c}{73.879(74)}              && \multicolumn{2}{c}{73.884(78)}              \\
$T$ (K)                                  & 6,999(245)        & 4,473(110)              && 7,015(245)        & 4,472(109)              && 6,997(245)        & 4,472(110)              \\
$\Omega$                                 & 4.799(32)         & 2.257                   && 4.816(37)         & 2.257                   && 4.799(37)         & 2.264                   \\
$A$                                      & 1.0               & 0.5                     && 1.0               & 0.5                     && 1.0               & 0.5                     \\
$g$                                      & 1.0               & 0.32                    && 1.0               & 0.32                    && 1.0               & 0.32                    \\
$X$, $Y$                                 & 0.639, 0.247      & 0.624, 0.160            && 0.639, 0.247      & 0.624, 0.160            && 0.639, 0.247      & 0.624, 0.160            \\
$x$, $y$                                 & 0.592, 0.293      & 0.737, 0.111            && 0.592, 0.293      & 0.737, 0.111            && 0.593, 0.292      & 0.737, 0.111            \\
$L$/($L_{1}$+$L_{2}$)                    & 0.8236(22)        & 0.1764                  && 0.8238(19)        & 0.1762                  && 0.8227(21)        & 0.1773                  \\
$r$ (pole)                               & 0.2177(18)        & 0.2361(12)              && 0.2169(20)        & 0.2360(13)              && 0.2178(20)        & 0.2370(15)              \\
$r$ (point)                              & 0.2199(18)        & 0.3458(17)              && 0.2191(21)        & 0.3457(18)              && 0.2201(21)        & 0.3469(20)              \\
$r$ (side)                               & 0.2190(18)        & 0.2456(13)              && 0.2182(21)        & 0.2456(14)              && 0.2192(21)        & 0.2465(15)              \\
$r$ (back)                               & 0.2197(18)        & 0.2780(13)              && 0.2189(21)        & 0.2780(14)              && 0.2198(21)        & 0.2790(16)              \\
$r$ (volume)$\rm ^a$                     & 0.2188(18)        & 0.2540(14)              && 0.2180(21)        & 0.2540(15)              && 0.2190(21)        & 0.2549(16)              \\ [1.0mm]
\multicolumn{6}{l}{Spot parameters:}                                                                                                                                                 \\        
Colatitude (deg)                         & \dots             & \dots                   && 45.5(4.9)         & \dots                   && \dots             & 83.4(5.6)               \\        
Longitude (deg)                          & \dots             & \dots                   && 94.5(1.5)         & \dots                   && \dots             & 64.3(1.5)               \\        
Radius (deg)                             & \dots             & \dots                   && 12.4(2.2)         & \dots                   && \dots             & 12.0(2.9)               \\        
$T$$\rm _{spot}$/$T$$\rm _{local}$       & \dots             & \dots                   && 1.041(10)         & \dots                   && \dots             & 0.896(46)               \\
$\sum$                                   & \multicolumn{2}{c}{0.00360}                 && \multicolumn{2}{c}{0.00317}                 && \multicolumn{2}{c}{0.00316}                 \\ [1.0mm]
\multicolumn{6}{l}{Absolute parameters:}                                                                                                                                             \\            
$M$ ($M_\odot$)                          & 1.50(8)           &  0.31(2)                && 1.50(8)           &  0.31(2)                && 1.50(8)           &  0.32(2)                \\
$R$ ($R_\odot$)                          & 1.51(4)           &  1.75(4)                && 1.50(4)           &  1.75(4)                && 1.51(4)           &  1.76(4)                \\
$\log$ $g$ (cgs)                         & 4.26(3)           &  3.45(4)                && 4.26(3)           &  3.45(4)                && 4.26(3)           &  3.45(4)                \\
$L$ ($L_\odot$)                          & 4.89(73)          &  1.10(12)               && 4.90(73)          &  1.10(12)               && 4.90(73)          &  1.11(12)               \\
$M_{\rm bol}$ (mag)                      & 3.01(16)          &  4.63(12)               && 3.00(16)          &  4.63(12)               && 3.00(16)          &  4.62(12)               \\
BC (mag)                                 & 0.03(1)           &  $-$0.62(8)             && 0.03(1)           &  $-$0.62(8)             && 0.03(1)           &  $-$0.62(8)             \\
$M_{\rm V}$ (mag)                        & 2.98(16)          &  5.25(15)               && 2.97(16)          &  5.25(14)               && 2.97(16)          &  5.24(15)               \\
\enddata
\tablenotetext{a}{Mean volume radius computed from the tables given by Mochnacki (1984). }
\end{deluxetable}

\begin{deluxetable}{lccc}
\tablewidth{0pt}
\tablecaption{Epoch and Spot Parameters for Each Light Curve of KIC 6048106 }
\tablehead{
\colhead{Epoch}    & \colhead{Longitude}  & \colhead{Angular Radius}   & \colhead{Temp. Factor}   \\
\colhead{(BJD)}    & \colhead{(deg)}      & \colhead{(deg)}            &                          
}
\startdata
2,456,108.55095    &  71.43               & 12.01                      & 0.896                    \\
2,456,110.11031    &  65.16               & 12.11                      & 0.919                    \\
2,456,111.66947    &  54.39               & 12.35                      & 0.899                    \\
2,456,113.22906    &  28.63               & 12.01                      & 0.913                    \\
2,456,114.78847    &  68.44               & 12.81                      & 0.927                    \\
2,456,116.34724    &  75.33               & 12.46                      & 0.924                    \\
2,456,117.90699    &  70.22               & 12.27                      & 0.914                    \\
2,456,119.46643    &  75.40               & 12.85                      & 0.942                    \\
2,456,121.02539    &  86.17               & 12.01                      & 0.939                    \\
2,456,122.58213    &  80.56               & 11.93                      & 0.924                    \\
\enddata
\tablecomments{This table is available in its entirety in machine-readable form. A portion is shown here for guidance regarding its form and content. }
\end{deluxetable}

\begin{deluxetable}{lrccccc}
\tablewidth{0pt}
\tablecaption{Multiple Frequency Analysis of KIC 6048106 }
\tablehead{
             & \colhead{Frequency}    & \colhead{Amplitude} & \colhead{Phase} & \colhead{S/N ratio} & \colhead{Remark}                \\
             & \colhead{(day$^{-1}$)} & \colhead{(mmag)}    & \colhead{(rad)} &                     &
}
\startdata                                                                                             
$f_{1}$      & 0.63775$\pm$0.00002    & 3.29$\pm$0.06       & 6.00$\pm$0.06   & 89.67               & $f_{\rm orb}$                   \\
$f_{2}$      & 1.97655$\pm$0.00003    & 1.77$\pm$0.05       & 4.58$\pm$0.09   & 56.23               &                                 \\
$f_{3}$      & 2.21165$\pm$0.00004    & 1.50$\pm$0.05       & 2.69$\pm$0.10   & 49.39               &                                 \\
$f_{4}$      & 2.04610$\pm$0.00004    & 1.39$\pm$0.05       & 2.81$\pm$0.11   & 44.55               &                                 \\
$f_{5}$      & 2.84140$\pm$0.00004    & 1.27$\pm$0.05       & 1.03$\pm$0.10   & 48.07               &                                 \\
$f_{6}$      & 2.30595$\pm$0.00005    & 1.07$\pm$0.05       & 5.53$\pm$0.14   & 35.92               &                                 \\
$f_{7}$      & 2.56510$\pm$0.00009    & 0.54$\pm$0.05       & 4.49$\pm$0.26   & 19.03               & 4$f_{\rm orb}$                  \\
$f_{8}$      & 3.20665$\pm$0.00004    & 1.04$\pm$0.04       & 1.57$\pm$0.12   & 43.37               & 5$f_{\rm orb}$                  \\
$f_{9}$      & 0.65200$\pm$0.00013    & 0.48$\pm$0.06       & 0.67$\pm$0.38   & 13.09               & $f_{\rm orb}$+0.01071           \\
$f_{10}$     & 2.72325$\pm$0.00008    & 0.60$\pm$0.05       & 6.04$\pm$0.23   & 22.02               &                                 \\
$f_{11}$     & 4.48925$\pm$0.00005    & 0.65$\pm$0.03       & 2.68$\pm$0.13   & 38.12               & 7$f_{\rm orb}$                  \\
$f_{12}$     & 1.28360$\pm$0.00007    & 0.85$\pm$0.06       & 1.97$\pm$0.20   & 24.66               & 2$f_{\rm orb}$                  \\
$f_{13}$     & 1.27415$\pm$0.00011    & 0.54$\pm$0.06       & 4.16$\pm$0.32   & 15.62               & 2$f_1$                          \\
$f_{14}$     & 0.64590$\pm$0.00011    & 0.60$\pm$0.06       & 1.90$\pm$0.31   & 16.37               & $f_{\rm orb}$                   \\
$f_{15}$     & 0.63005$\pm$0.00012    & 0.53$\pm$0.06       & 5.69$\pm$0.35   & 14.36               & $f_{\rm orb}-$0.01124           \\
$f_{16}$     & 3.03275$\pm$0.00010    & 0.44$\pm$0.04       & 3.06$\pm$0.29   & 17.51               & 7$f_{\rm orb}$+2$f_1$-$f_{10}$  \\
$f_{17}$     & 2.79950$\pm$0.00017    & 0.27$\pm$0.05       & 3.93$\pm$0.50   & 10.09               & 4$f_{\rm orb}-$$f_2$+$f_3$      \\
$f_{18}$     & 1.30860$\pm$0.00025    & 0.24$\pm$0.06       & 3.32$\pm$0.72   &  6.94               & 2$f_9$                          \\
$f_{19}$     & 1.27985$\pm$0.00019    & 0.32$\pm$0.06       & 6.17$\pm$0.54   &  9.32               & 2$f_{\rm orb}$                  \\
$f_{20}$     & 5.77215$\pm$0.00007    & 0.29$\pm$0.02       & 4.67$\pm$0.22   & 23.11               & 9$f_{\rm orb}$                  \\
$f_{21}$     & 2.82475$\pm$0.00020    & 0.23$\pm$0.05       & 6.25$\pm$0.57   &  8.75               & 8$f_{\rm orb}-$$f_6$            \\
$f_{22}$     & 2.41375$\pm$0.00024    & 0.21$\pm$0.05       & 3.72$\pm$0.69   &  7.31               & 5$f_{\rm orb}$+$f_4-$$f_5$      \\
$f_{23}$     & 1.85800$\pm$0.00026    & 0.22$\pm$0.06       & 6.07$\pm$0.74   &  6.75               & $f_2-$$f_5$+$f_{10}$            \\
$f_{24}$     & 1.26435$\pm$0.00030    & 0.20$\pm$0.06       & 4.48$\pm$0.85   &  5.87               & 2$f_{15}$                       \\
$f_{25}$     & 0.31565$\pm$0.00032    & 0.20$\pm$0.06       & 2.68$\pm$0.94   &  5.34               & 7$f_{\rm orb}$+2$f_1-$2$f_{10}$ \\
$f_{26}$     & 2.68710$\pm$0.00026    & 0.18$\pm$0.05       & 2.33$\pm$0.75   &  6.64               & $f_{\rm orb}$+$f_4$             \\
$f_{27}$     & 1.11995$\pm$0.00033    & 0.19$\pm$0.06       & 2.19$\pm$0.95   &  5.30               & 6$f_{\rm orb}-$$f_{10}$         \\
$f_{28}$     & 2.65335$\pm$0.00029    & 0.17$\pm$0.05       & 2.31$\pm$0.83   &  6.06               & $f_2$-$f_4$+$f_{10}$            \\
$f_{29}$     & 2.11240$\pm$0.00032    & 0.17$\pm$0.05       & 1.64$\pm$0.92   &  5.45               & 2$f_4-$$f_2$                    \\
$f_{30}$     & 1.69665$\pm$0.00042    & 0.14$\pm$0.06       & 4.10$\pm$1.21   &  4.15               & 3$f_{\rm orb}$+$f_2-$$f_3$      \\
\enddata                                                                                                                           
\end{deluxetable}

\begin{deluxetable}{lccccccc}
\tabletypesize{\small}
\tablewidth{0pt}
\tablecaption{$\gamma$ Dor-type Pulsation Properties of KIC 6048106 }
\tablehead{
       & \colhead{Frequency}    & \colhead{$Q$}    & \colhead{($f_i$/$f_5$)$_{\rm obs}$} & \colhead{mode ($n$, $\ell$)}  & \colhead{($f_i$/$f_5$)$_{\rm model}$} & \colhead{$\Delta$($f_i$/$f_5$)$_{\rm obs-model}$} & \colhead{$\cal J_{\rm obs}$}    \\
       & \colhead{(day$^{-1}$)} & \colhead{(days)} &                                     &                               &                                       &                                                   & \colhead{($\mu$Hz)}  
}
\startdata		
$f_{2}$  & 1.97655   & 0.333  & 0.6956   & (25, 2)    & 0.6863   & +0.0094               & 748.2            \\
$f_{3}$  & 2.21165   & 0.298  & 0.7784   & (22, 2)    & 0.7778   & +0.0006               & 738.7            \\
$f_{4}$  & 2.04610   & 0.322  & 0.7201   & (24, 2)    & 0.7143   & +0.0058               & 744.1            \\
$f_{5}$  & 2.84140   & 0.232  & \dots    & (17, 2)    & \dots    & \dots                 & 738.1            \\
$f_{6}$  & 2.30595   & 0.286  & 0.8116   & (21, 2)    & 0.8140   & $-$0.0024             & 735.9            \\
$f_{10}$ & 2.72325   & 0.242  & 0.9584   & (18, 2)    & 0.9459   & +0.0125               & 747.9            \\
         &           &        &          &            & Average  & $+$0.0052$\pm$0.0061  & 742.2$\pm$5.3    \\  
\enddata                                                                                                                           
\end{deluxetable}


\newpage
\begin{thebibliography}{}
\bibitem[Balona et al(2011)]{balona2011} Balona, L. A., Guzik, J. A., Uytterhoeven, K., et al. 2011, MNRAS, 415, 3531
\bibitem[Bradley et al(2015)]{bradley2015} Bradley, P. A., Guzik, J. A., Miles, L. F., et al. 2015, AJ, 149, 68
\bibitem[Breger et al(1993)]{breger1993} Breger, M., Stich, J., Garrido, R., et al. 1993, A\&A, 271, 482
\bibitem[Cakirli(2015)]{cakirli2015} \c Cakirli, \" O. 2015, New Astron., 38, 55
\bibitem[Cakirli \& Ibano\v{g}lu(2016)]{cakirli2016} \c Cakirli, \" O., \& Ibano\v{g}lu, C. 2016, New Astron., 45, 36
\bibitem[Dupret et al(2004)]{dupret2004} Dupret, M.-A., Grigahc\'ene, A., Garrido, R., Gabriel, M., \& Scuflaire, R. 2004, A\&A, 414, 17
\bibitem[Dupret et al(2005)]{dupret2005} Dupret, M.-A., Grigahc\'ene, A., Garrido, R., Gabriel, M., \& Scuflaire, R. 2005, A\&A, 435, 927
\bibitem[Gaulme et al(2014)]{gaulme2014} Gaulme, P., \& Guzik, J. A. 2014, in Proc. IAU Symp. 301, Precision Asteroseismology, ed. J. A. Guzik, W. Chaplin, G. Handler, \& A. Pigulski (Cambridge: Cambridge Univ. Press), 413
\bibitem[Guzik et al(2000)]{guzik2000} Guzik, J. A., Kaye, A. B., Bradley, P. A., Cox, A. N., \& Neuforge C. 2000, ApJ, 542, L57
\bibitem[Hambleton et al(2013)]{hambleton2013} Hambleton, K. M., Kurtz, D. W., Pr\v sa, A. et al. 2013, MNRAS, 434, 925
\bibitem[Hareter et al(2010)]{hareter2010} Hareter, M., Reegen, P., Miglio, A., et al. 2010, arXiv:1007.3176
\bibitem[Harmanec(1988)]{harmanec1988} Harmanec, P. 1988, Bull. Astron. Inst. Czechoslovakia, 39, 329
\bibitem[Henry(2005)]{henry2005} Henry, G. W., Fekel, F. C., \& Henry, S. M. 2005, AJ, 129, 2815
\bibitem[Huber et al(2014)]{huber2015} Huber, D., Silva Aguirre, V., Matthews, J. M., et al. 2014, ApJS, 211, 2
\bibitem[Ibano\v{g}lu et al(2006)]{ibanoglu2006} \. Ibano\v{g}lu, C., Soydugan, F., Soydugan, E., \& Dervi\c so\v{g}lu, A. 2006, MNRAS, 373, 435
\bibitem[Kallinger et al(2008)]{kallinger2008} Kallinger, T., Reegen, P., \& Weiss, W. W. 2008, A\&A, 481, 571
\bibitem[Kaye et al(1999)]{kaye1999} Kaye, A. B., Handler, G., Krisciunas, K., Poretti, E., \& Zerbi, F. M. 1999, PASP, 111, 840
\bibitem[Koo et al(2014)]{koo2014} Koo, J.-R., Lee, J. W., Lee, B.-C., et al. 2014, AJ, 147, 104
\bibitem[Kurtz et al(2015)]{kurtz2015} Kurtz, D. W., Hambleton, K. M., Shibahashi, H., Murphy, S. J., \& Pr\v sa, A. 2015, MNRAS, 446, 1223
\bibitem[Lee et al(2016a)]{lee2016a} Lee, J. W., Hong, K., Kim, S.-L., \& Koo, J.-R. 2016a, MNRAS, 460, 4220
\bibitem[Lee et al(2016b)]{lee2016b} Lee, J. W., Kim, S.-L., Hong, K., Koo, J.-R., Lee, C.-U., \& Youn, J.-H. 2016b, AJ, 151, 25
\bibitem[Lee et al(2014)]{lee2014} Lee, J. W., Kim, S.-L., Hong, K., Lee, C.-U., \& Koo, J.-R. 2014, AJ, 148, 37
\bibitem[Lenz \& Breger(2005)]{len2005} Lenz, P., \& Breger, M. 2005, Comm. Asteroseismology, 146, 53
\bibitem[Liakos et al(2012)]{liakos2012} Liakos, A., Niarchos, P., Soydugan, E., \& Zasche, P. 2012, MNRAS, 422, 1250
\bibitem[Liu et al(2012)]{liu2012} Liu, N., Zhang, X. B., Ren, A. B., Deng, L. C., \& Luo, Z. Q. 2012, Res. Astron. Astrophys., 12, 671
\bibitem[Loumos \& Deeming(1978)]{loumos1978} Loumos, G. L., \& Deeming T. J. 1978, Ap\&SS, 56, 285
\bibitem[Maceroni et al(2014)]{maceroni2014} Maceroni, C., Lehmann, H., da Silva, R., et al. 2014, A\&A, 563, 59
\bibitem[Mkrtichian et al(2004)]{mkrtichian2004} Mkrtichian, D. E., Kusakin, A. V., Rodriguez, E., et al. 2004, A\&A, 419, 1015
\bibitem[Mochnacki(1984)]{mochnacki1984} Mochnacki, S. W. 1984, ApJS, 55, 551
\bibitem[Moya et al(2005)]{Moya2005} Moya, A., Su\'arez, J. C., Amado, P. J., et al. 2005, A\&A, 432, 189
\bibitem[Petersen \& Jorgensen(1972)]{petersen1972} Petersen, J. O., \& J\o rgensen, H. E. 1972, A\&A, 17, 367
\bibitem[Prsa et al(2011)]{prsa2011} Pr\v sa, A., Batalha, N., Slawson, R. W., et al. 2011, AJ, 141, 83
\bibitem[Rolland et al(2002)]{rolland2002} Rolland, A., Costa, V., Rodr\'iguez, E., et al. 2002, Comm. Asteroseismology, 142, 57
\bibitem[Slawson et al(2011)]{slawson2011} Slawson, R. W., Pr\v sa, A., Welsh, W. F., et al. 2011, AJ, 142, 160
\bibitem[Soydugan et al(2006)]{soydugan2006} Soydugan, E., \. Ibano\v{g}lu, C., Soydugan, F., Akan, M. C., \& Demircan, O. 2006, MNRAS, 366, 1298
\bibitem[Suarez et al(2005)]{Suarez2005} Su\'arez, J. C., Moya, A., Mart\'in-Ru\'iz, S., Amado, P. J., Grigahc\`ene, A., \& Garrido, R. 2005, A\&A, 443, 271
\bibitem[Torres(2010)]{torres2010} Torres, G. 2010, AJ, 140, 1158
\bibitem[Tout(1996)]{tout1996} Tout, C. A., Pols, O. R., Eggleton, P. P., \& Han, Z. 1996, MNRAS, 281, 257
\bibitem[Murphy(2012)]{murphy2012} Murphy, S. J., 2012, MNRAS, 422, 665
\bibitem[Murphy et al(2013)]{murphy2013} Murphy, S. J., Shibahashi, H., \& Kurtz, D. W., 2013, MNRAS, 430, 2986
\bibitem[Van Hamme \& Wilson(2007)]{van2007} Van Hamme, W., \& Wilson, R. E. 2007, ApJ, 661, 1129
\bibitem[Warner et al(2003)]{warner2003} Warner, P. B., Kaye, A. B., \& Guzik, J. A. 2003, ApJ, 593, 1049
\bibitem[Welsh et al(2011)]{welsh2011} Welsh, W. F., Orosz, J. A., Aerts, C., et al. 2011, ApJS, 197, 4
\bibitem[Wilson \& Devinney(1971)]{wilson1971} Wilson, R. E., \& Devinney, E. J. 1971, ApJ, 166, 605
\end{thebibliography}
\end{document}